\documentclass[pra, aps, twocolumn, groupedaddress, showpacs, superscriptaddress]{revtex4-2}
\usepackage{amssymb,amsmath,amsthm,color,graphicx,times,graphicx}
\usepackage{hyperref}
\usepackage{subfigure}
\usepackage{times}
\usepackage{bbold}
\usepackage{color}

\newcommand{\bra}[1]{\langle{#1}|}
\newcommand{\ket}[1]{|{#1}\rangle}

\providecommand{\openone}{\leavevmode\hbox{\small1\kern-4.3pt\normalsize1}}
\renewcommand\Re{\mathrm{Re}}

\renewcommand{\Re}{\mathrm{Re}}
\renewcommand{\Im}{\mathrm{Im}}

\theoremstyle{plain}

\theoremstyle{definition}

\usepackage{orcidlink}
\makeatletter
\newsavebox{\@brx}
\newcommand{\llangle}[1][]{\savebox{\@brx}{\(\m@th{#1\langle}\)}%
  \mathopen{\copy\@brx\mkern2mu\kern-0.9\wd\@brx\usebox{\@brx}}}
\newcommand{\rrangle}[1][]{\savebox{\@brx}{\(\m@th{#1\rangle}\)}%
  \mathclose{\copy\@brx\mkern2mu\kern-0.9\wd\@brx\usebox{\@brx}}}
\makeatother
\begin{document}
\title{Characterizing non-Markovianity via quantum coherence based on Kirkwood-Dirac quasiprobability}
\author{Yassine Dakir \orcidlink{0009-0005-3408-1309}}\affiliation{LPHE-Modeling and Simulation, Faculty of Sciences, Mohammed V University in Rabat, Rabat, Morocco.}
\author{Abdallah Slaoui \orcidlink{0000-0002-5284-3240}}\email{Corresponding author: abdallah.slaoui@um5s.net.ma}\affiliation{LPHE-Modeling and Simulation, Faculty of Sciences, Mohammed V University in Rabat, Rabat, Morocco.}\affiliation{Centre of Physics and Mathematics, CPM, Faculty of Sciences, Mohammed V University in Rabat, Rabat, Morocco.}\affiliation{Center of Excellence in Quantum and Intelligent Computing, Prince Sultan University, Riyadh 11586, Saudi Arabia.}
\author{Rachid Ahl Laamara\orcidlink{0000-0002-8254-9085}}\affiliation{LPHE-Modeling and Simulation, Faculty of Sciences, Mohammed V University in Rabat, Rabat, Morocco.}\affiliation{Centre of Physics and Mathematics, CPM, Faculty of Sciences, Mohammed V University in Rabat, Rabat, Morocco.}

\begin{abstract}
We present a new measure of non-Markovianity based on the property of nonincreasing quantum coherence via Kirkwood-Dirac (KD) quasiprobability under incoherent completely positive trace-preserving maps. Quantum coherence via the KD quasiprobability is defined as the imaginary part of the KD quasiprobability, which is maximised over all possible second bases and evaluated using an incoherent reference basis. A measure non-Markovianity based on KD quasiprobability coherence would capture memory effects via the time evolution of the imaginary part of the KD quasiprobability, providing an experimentally accessible and physically intuitive alternative to traditional measures relying on quantum Fisher information or trace distance. This approach is applied to the study of dissipation and dephasing dynamics in single- and two-qubit systems. The results obtained show that, in the cases studied, our measure based on coherence via Kirkwood-Dirac quasiprobability performs at least as well as $\ell_{1}$-norm coherence in detecting non-Markovianity, this provides a novel perspective on the analysis of non-Markovian dynamics.\par

\vspace{0.25cm}
\textbf{Keywords}: Non-Markovianity, quantum coherence, Kirkwood-Dirac quasiprobability.
%\pacs{03.65.Ta, 03.65.Yz, 03.67.Mn, 42.50.-p, 03.65.Ud}
\end{abstract}
\date{\today}

\maketitle
\section{Introduction}
In recent years, open quantum systems with memory is a rapidly expanding field of research, driven by their ability to model a wide range of physical phenomena and provide new insights into the control of quantum properties. These systems emerge from interactions between a system and its environment, whose dynamics fall into two principal categories: Markovian and non-Markovian \cite{Breuer2002,Banerjee2018,Candeloro2021,Dakir2023}. In Markovian dynamics, information is irreversibly transferred from the system to the environment without feedback or memory effects (i.e., a memoryless environment) \cite{Breuer2009,Rivas2010,Benedetti2014,Bylicka2017}. Conversely, non-Markovian dynamics involve memory effects, marked by a backflow of information from the environment to the system. Extensive research has highlighted the significance of non-Markovian structured environments in various fields \cite{Huelga2012,Chin2012,Vasile2011,Bouhouch2025,Laine2014,Slaoui2023,Slaoui2024,Abouelkhir2024,Thorwart2009}, including complex state generation, quantum metrology, quantum key distribution, quantum teleportation, and quantum biology, underscoring their potential advantages over Markovian channels.\par

Motivated by these diverse considerations, researchers have proposed multiple indicators and measures of non-Markovianity.\cite{Rivas2014,Luo2012,Rajagopal2010,Chen2016,Lu2010,Song2015,Abiuso2023,He2014,Chanda2016,He2017,Frigerio2021,Dakir2024}. These approaches rely on a range of considerations, including divisibility, quantum mutual information, fidelity, temporal steering, informational distance measures, quantum Fisher information, local quantum uncertainty, relative entropy, $\ell_{1}$-norm coherence, and local quantum Fisher information, as well as practical experimental methods \cite{Liu2011,Tang2012,Bernardes2015}, among others. However, these measures do not always align perfectly when quantifying non-Markovian processes, though they may converge in specific cases. The study of non-Markovianity measures is now attracting growing interest both theoretically and experimentally. It has been shown that, in the case of open dynamics of single channels (characterized by a single decay rate), these different measures are equivalent \cite{Addis2014}. However, in more complex scenarios, they generally diverge. Establishing a hierarchy among these measures and developing a general definition of non-Markovianity continue to pose significant challenges in current research.\par

Quantum coherence (QC), which comes from quantum superposition, is a key part of quantum mechanics and is behind many of the special and interesting properties of quantum systems. One of those things is what makes the quantum world different from the classical world \cite{Leggett1980}. This property has profound implications in various fields, including quantum optics \cite{Zhang012,Walls1995}, solid-state physics \cite{Deveaud2009,Li2012}, and quantum thermodynamics \cite{Robnagel2014,Tarif2025,Hminat2025}. Despite being recognized as a valuable physical resource, the precise quantification of quantum coherence remains a major challenge. Recent advances have yielded a systematic approach to coherence quantification \cite{Baumgratz2014,Rana2016,Shao2015,Yu2017,Rastegin2016,Streltsov2015,Hu2018,Streltsov2017}, employing multiple metrics such as, relative entropy of coherence, $\ell_{1}$-norm coherence, fidelity-based coherence, and skew information coherence measures. Satisfying monotonicity under all completely positive, trace-preserving incoherent maps (CPTP), these coherence measures have emerged as powerful tools for characterizing quantum dynamics, especially in identifying non-Markovian effects. The $\ell_{1}$-norm coherence-based measure proposed by Chanda et al. \cite{Chanda2016} and the relative entropy approach of He et al.\cite{He2017} successfully identify non-Markovianity, showing that they are similar to measures based on quantum divisibility and trace distance.\par

In this work, we introduce a new way to measure non-Markovianity for incoherent dynamical maps. Our new measure is based on quantum coherence via KD quasiprobability.  Our method quantifies the nonmonotonic behavior of quantum coherence, defined by the imaginary part of the KD quasiprobability. We evaluate this measure against two baselines, an incoherent reference and an optimized baseline considering all possible choices. Our approach enables efficient detection of non-Markovianity without requiring auxiliary systems, making it experimentally feasible even in the context of complex quantum dynamics.\par

This work is structured as follows: Sec. (\ref{sec2}) presents the theoretical foundations of concepts of quantum coherence via the Kirkwood-Dirac quasiprobability. In Sec. (\ref{sec3}), we suggest a way of detecting non-Markovianity based on quantum coherence via the KD quasiprobability. In Sec.(\ref{sec4}), we demonstrate the effectiveness of our proposed measure through several illustrative examples. Finally, Sec.(\ref{clc}) summarizes our findings and discusses future directions.

\section{quantum coherence based on KD quasiprobability} \label{sec2}
We consider a finite-dimensional Hilbert space (HS) $\mathcal{H}$ with dimension $d = \dim(\mathcal{H})$. An orthonormal basis $\left\lbrace \ket{\mu} \right\rbrace$ is fixed and the associated projectors $\mathcal{X}_{\mu} = \ket{\mu} \bra{\mu}$ satisfy $\sum_{\mu} \mathcal{X}_{\mu} = \mathbb{I}$, where $\mathbb{I}$ is the identity operator. Any diagonal density matrix in this basis is called an incoherent state. A general incoherent state is indicated by $\sigma = \sum_{\mu} \vartheta_{\mu} \ket{\mu}\bra{\mu}$, where $\vartheta_{\mu} \geq 0$ and $\sum_{\mu} \vartheta_{\mu} = 1$, the set of incoherent states, denoted $\mathcal{I}_{c}$, is thus defined as the subset of diagonal density matrices in the space $\mathcal{H}$.\par

In this context, Budiyono and Dipojono proposed a novel measure for detecting quantum coherence, derived from the imaginary part of the KD quasiprobability \cite{Budiyono2023}. In a finite-dimensional HS, an intrinsic non-commutativity of quantum mechanics allows the KD quasiprobability to take complex or negative values, making it an  informatively equivalent representation of a quantum state. For a HS $\mathcal{H}$ of dimension $d$, two orthonormal bases $\left\lbrace \ket{\mu} \right\rbrace$ and $\left\lbrace \ket{\nu} \right\rbrace$, and a quantum state $\varrho$ defined on this space, the KD quasiprobability is given by 
\begin{equation}
    \mathcal{P}_{KD}\left(\mu,\nu|\varrho\right) = Tr({\mathcal{X}_{\nu}\mathcal{X}_{\mu}\varrho}) \label{P}
\end{equation}
It is simple to confirm that the KD quasiprobability may properly describe the quantum state $\varrho$ as
\begin{equation}
    \varrho=\sum_{\mu,\nu} \bra{\mu} \varrho\ket{\nu}\ket{\mu}\bra{\nu}=\sum_{\mu,\nu} \mathcal{P}_{KD}(\mu,\nu|\varrho)\frac{\ket{\mu}\bra{\nu}}{<\mu\arrowvert \nu>}
\end{equation}
The KD quasiprobability gives correct marginal probabilities 
\begin{align}
    &\sum_{\mu}\mathcal{P}_{KD}(\mu,\nu|\varrho)=Tr(\mathcal{X}_{\nu}\varrho)\\
    &\sum_{\nu}\mathcal{P}_{KD}(\mu,\nu|\varrho)=Tr(\mathcal{X}_{\mu}\varrho)
\end{align}
and thus normalized $\sum_{\mu,\nu}\mathcal{P}_{KD}(\mu,\nu|\varrho)=Tr(\mathcal{X}_{\nu}\varrho)=1$. A quantum state can be represented in a complete way using the KD quasiprobability distribution. This naturally raises the question of how quantum coherence of a state with respect to a given incoherent basis is encoded in this distribution. While quantum coherence is conventionally characterized relative to a single fixed incoherent basis, the Kirkwood-Dirac (KD) quasiprobability is fundamentally defined through two distinct reference bases. To address this question, we leverage a key property demonstrated in previous studies: the imaginary component of the KD quasiprobability directly captures the commutator between the quantum state $\rho$ and any chosen basis $\{\ket{\mu}\}$. This observation establishes a direct link between the mathematical structure of KD quasiprobability and the representation of quantum coherence, the quantum state and an incoherent reference basis $\left\lbrace\ket{\mu}\right\rbrace$. On this basis, we define a quantity that associates a non-negative real number with the quantum state
\begin{align}
    \mathcal{C}_{KD}[\varrho; \left\lbrace \mathcal{X}_{\mu}\right\rbrace] &= \max_{\left\lbrace \ket{\mu}\right\rbrace}\sum_{\mu}\sum_{\nu}\arrowvert \Im \left\lbrace \mathcal{P}_{KD}(\mu,\nu|\varrho)\right\rbrace\arrowvert \notag\\
    &= \max _{\left\lbrace \ket{\mu}\right\rbrace}\sum_{\mu}\sum_{\nu}\arrowvert \Im \left\lbrace \bra{\nu}\mathcal{X}_{\nu}\varrho\ket{\nu}\arrowvert\right\rbrace \notag\\
    &=
    \max _{\left\lbrace \ket{\mu}\right\rbrace}\sum_{\mu}\sum_{\nu}\arrowvert \frac{1}{2} \arrowvert \bra{\nu}[\mathcal{X}_{\mu}, \varrho]\ket{\mu}\arrowvert \label{ckd}
\end{align}
where $\{\ket{\nu}\}$ represents another basis of $\mathcal{H}_S$. To quantify this relationship, we compute the $\ell_{1}$-norm of the imaginary part of the KD quasiprobability $\mathcal{P}_{KD}(\mu,\nu|\varrho)$, maximizing it over all possible choices of the second basis $\{\ket{\nu}\}$. This maximization identifies the maximal quantum coherence between the quantum state $\varrho$ and the reference basis $\{\ket{\mu}\}$, as revealed through measurements in the $\{\ket{\nu}\}$ basis using the $\ell_{1}$-norm. Therefore, a valid $\mathcal{C}_{KD}$ coherence measure of state $\varrho$ should satisfy the following properties:

\begin{enumerate}
\item[(\textbf{A1})] \textbf{Faithful} :\\
The quantum state $\varrho$ is incoherent with respect to the basis $\left\lbrace \ket{\mu}\right\rbrace$ if and only if:
\begin{equation}
    \mathcal{C}_{KD}[\varrho,\left\lbrace \mathcal{X}_{\mu}\right\rbrace]=0
\end{equation}
    \item[(\textbf{A2})] \textbf{Convexity:} 
\begin{equation}
    \mathcal{C}_{KD}\left[\sum_{n}\vartheta_{j}\varrho_{j}; \left\lbrace \mathcal{X}_{\mu}\right\rbrace\right] \geq \sum_{j} \vartheta_{j} \mathcal{C}_{KD}\left[\varrho_{j}; \left\lbrace \mathcal{X}_{\mu}\right\rbrace\right],
\end{equation}
where $\vartheta_{j}$ represents probabilities, satisfying $0 \leq \vartheta_{j} \leq 1$ and $\sum_{j} \vartheta_{j} = 1$.

\item[(\textbf{A3})] \textbf{Unitary covariance} : 

\begin{equation}
        \mathcal{C}_{KD}[U\varrho U^{\dagger}; \left\lbrace U \mathcal{X}_{\mu} U^{\dagger}\right\rbrace] = \mathcal{C}_{KD}[\varrho; \left\lbrace \mathcal{X}_{\mu}\right\rbrace]
\end{equation}

\item[(\textbf{A4})] \textbf{Non-increasing under partial trace} : 
\begin{equation}
    \mathcal{C}_{KD}\left[\varrho_{a}; \left\lbrace \mathcal{X}_{\mu_{a}} \otimes \mathcal{I}_{b}\right\rbrace\right] \leq \mathcal{C}_{KD}\left[\varrho_{ab}; \left\lbrace \mathcal{X}_{\mu_{a}}\right\rbrace\right],
\end{equation}
where $\varrho_{ab}$ is the quantum state of the composite system of subsystems $a$ and $b$, $\varrho_{a} = Tr_{b}(\varrho_{ab})$ is the reduced state of subsystem $a$, and $\mathcal{I}_{b}$ is the identity operator of subsystem $b$.

\item[(\textbf{A5})] \textbf{Monotonicity under ICPTP maps} : 
\begin{equation}
    \mathcal{C}_{KD}[\Phi(\varrho); \left\lbrace \mathcal{X}_{\mu}\right\rbrace] \leq \mathcal{C}_{KD}[\varrho; \left\lbrace \mathcal{X}_{\mu}\right\rbrace],
\end{equation}
where $\Phi$ is a completely positive, trace-preserving incoherent map (ICPTP).
\end{enumerate}
\textit{Proof of property (A5):}\\
First, when $\Phi$ is a random unitary channel, there is
\begin{equation}
    \Phi(\varrho) = \sum_{k} p_{k}U_{k}\varrho U_{k}^{\dagger}
\end{equation}
where $U_{k}$ are unitary operators on $H$, and $p_{k}$ satisfy
$0\leq p_{k} \leq 1$, $\sum_{k}p_{k}=1$. Then we have
\begin{align}
    \Im \left\lbrace Tr[\Phi(\varrho) \mathcal{X}_{\mu} \ket{\nu}\bra{\nu}]\right\rbrace&= \Im \left\lbrace Tr[(\sum_{k} p_{k}U_{k}\varrho U_{k}^{\dagger}) \mathcal{X}_{\mu} \ket{\nu}\bra{\nu}]\right\rbrace \notag\\
    &=\sum_{k}p_{k}\Im \left\lbrace Tr[U_{k}\varrho U_{k}^{\dagger} \mathcal{X}_{\mu} \ket{\nu}\bra{\nu}]\right\rbrace \notag\\
    &=\sum_{k}p_{k}\Im \left\lbrace Tr[\varrho U_{k}^{\dagger}\mathcal{X}_{\mu}U_{k}  \ket{\nu}\bra{\nu}]\right\rbrace
\end{align}
So, 
\begin{align}
    \mathcal{C}_{KD}[\Phi(\varrho), \left\lbrace \mathcal{X}_{\mu}\right\rbrace]&=\max_{\left\lbrace \ket{\nu}\right\rbrace} \sum_{\mu,\nu}\arrowvert \sum_{k}p_{k}\Im \left\lbrace Tr[U_{k}\varrho U_{k}^{\dagger} \mathcal{X}_{\mu} \ket{\nu}\bra{\nu}]\right\rbrace\arrowvert \notag\\
    &\leq \max_{\left\lbrace \ket{\nu}\right\rbrace}\sum_{\mu,\nu}\sum_{k}p_{k}\arrowvert\Im \left\lbrace Tr[\varrho U_{k}^{\dagger}\mathcal{X}_{\mu}U_{k}  \ket{\nu}\bra{\nu}]\right\rbrace \arrowvert \notag\\
    &=\max_{\left\lbrace \ket{\nu}\right\rbrace}\sum_{\mu,\nu}\sum_{k}p_{k}\arrowvert\Im \left\lbrace Tr[\varrho\mathcal{X}_{\mu}  \ket{\nu}\bra{\nu}]\right\rbrace \arrowvert \notag\\ 
    &=\mathcal{C}_{KD}[\varrho, \left\lbrace \mathcal{X}_{\mu}\right\rbrace] \label{NC}
\end{align}

Next, from the decomposition of KD quasiprobability given by Eq. (\ref{P}), we quantifies KD nonclassicality, i.e., the negativity and the nonreality in the KD quasiprobability, with a quantity $N_{c}[\mu,\nu|\varrho]$ of the density matrix of $\varrho$ and $\mathcal{X}_{\nu_{i}}$ under $\left\lbrace \mathcal{X}_{\mu}\right\rbrace$ defined as \cite{He2024}
\begin{align}
    N_{c}[\mu,\nu|\varrho]=\sum_{i,j}&\Bigg[ |\sum_{k \neq i}\Re \left\lbrace \bra{\mu_{i}}\varrho\ket{\mu_{k}}\bra{\mu_{k}}\mathcal{X}_{\nu_{i}}\ket{\mu_{i}}\right\rbrace| \notag \\
    & +|\sum_{k \neq i}\Im \left\lbrace \bra{\mu_{i}}\varrho\ket{\mu_{k}}\bra{\mu_{k}}\mathcal{X}_{\nu_{i}}\ket{\mu_{i}}\right\rbrace|\Bigg] \label{nonclass}
\end{align} 
\section{Quantifying non-Markovianity via Kirkwood-Dirac quasiprobability} \label{sec3}
We consider quantum processes evolving according to the time-local master equation,
\begin{equation}
    \frac{d\varrho(t)}{dt}=\mathcal{D}_{t}[\varrho(t)]
\end{equation}
where $\mathcal{D}_{t}$ is a superoperator acting on the reduced density matrix $\varrho(t)$ is given by \cite{Gorini1976,Lindblad1976}
\begin{align}
    \mathcal{D}_{t}[\varrho(t)]=-i[H(t),\varrho(t)]&-\sum_{k}\gamma_{k}(t)\Bigg(\mathcal{A}_{k}(t)\varrho(t)\mathcal{A}_{k}^{\dagger}(t) \notag\\
    -&\frac{1}{2} [\mathcal{A}_{k}^{\dagger}(t)\mathcal{A}_{k}(t),\varrho(t)]_{+}\bigg)
\end{align}
where $[X,Y]_{+}$ defines the anti-commutator of $X$ and $Y$, $\gamma_{k}(t)$ are the time dependent relaxation rate and $\mathcal{A}_{k}(t)$ are the Lindblad operators.

The quantum evolution is considered Markovian if $\gamma_{k}(t) \geq 0$ at all times $(\forall t \geq 0)$.The dynamics of the system may be characterized via time-ordered CPTP maps. These maps are given by the expression 
\begin{equation}
    \Phi(t_{2},t_{1}) = V \exp\left[\int_{t_{1}}^{t_{2}}  \mathcal{D}_{t'}dt'\right],
\end{equation}
where $V$ is the time ordering operator. The map $\Phi(t,0)$ captures the dynamics of the system from its initial state at $t = 0$ to the state it is in at a specified time $t$. A dynamical process is classified as Markovian if the CPTP map satisfies the divisibility condition, 
\begin{equation}
    \Phi(t) = \Phi(t,\upsilon)\Phi(\upsilon), \quad \forall t \geq 0 \text{ and } \upsilon \geq 0.
\end{equation}
However, when $\gamma_{k}(t)$ becomes negative at any point, the corresponding master equation describes non-Markovian dynamics, indicating a deviation from Markovian behavior. As a measure of quantum coherence based of KD quasiprobability, $\mathcal{C}_{KD}\left[ \varrho\left( t\right)\right]$ satisfies the monotonicity under incoherent CPTP maps, i.e $\mathcal{C}_{KD}\left[ \varrho\left( t\right), \left\lbrace \mathcal{X}_{\mu}\right\rbrace\right]\leq \mathcal{C}_{KD}\left[\varrho, \left\lbrace \mathcal{X}_{\mu}\right\rbrace\right]$. \\

\textit{Proof of the monotonicity of CKD quasiprobability under Markovian dynamics:}\\
For a quantum state $\varrho(t)$ at time $t$ is determined by $\varrho(t) = \Phi_{t} \varrho(0)$ for a quantum system whose initial state $\varrho(0)$ undergoes an incoherent dynamic evolution CPTP $\Phi_{t}$ maps. Consequently, the CKD quasiprobability of the compound system satisfies   
\begin{align}
	\mathcal{C}_{KD}\left[\varrho\left( t\right), \left\lbrace \mathcal{X}_{\mu}\right\rbrace\right]&=\mathcal{C}_{KD}\left( \Phi_{t} \varrho(0), \left\lbrace \mathcal{X}_{\mu}\right\rbrace\right] \notag\\
	&= \mathcal{C}_{KD}\left[ \Phi_{t,\upsilon}\Phi_{\upsilon} \varrho(0), \left\lbrace \mathcal{X}_{a}\right\rbrace\right] \notag\\
	&=\mathcal{C}_{KD}\left[ \Phi_{t,\upsilon} \varrho\left( \upsilon\right), \left\lbrace \mathcal{X}_{\mu}\right\rbrace\right] \notag\\
	&\leq \mathcal{C}_{KD}\left[ \varrho\left( \upsilon\right), \left\lbrace \mathcal{X}_{\mu}\right\rbrace\right].
\end{align}
Together with the property of divisibility, one can conclude that the CKD quasiprobability decreases monotonically under Markovian dynamics and violations of this monotonicity indicate the occurrence of non-Markovian dynamics. Since the CKD quasiprobability is monotonically non-increasing under incoherent CPTP maps, the function $\mathcal{C}_{KD}\left[ \varrho\left( t\right)\right]$ is monotonically decreasing of $t \leq 0$. Hence,
\begin{equation}
    \frac{d\mathcal{C}_{KD}\left[ \varrho\left( t\right)\right]}{dt} \leq 0
\end{equation}
Therefore, our measure of non-Markovianity is defined by integrating the positive slope of the CKD quasiprobability $(\sigma_{\mathcal{C}}(t)=\frac{d\mathcal{C}_{KD}\left[ \varrho\left( t\right)\right]}{dt} > 0)$. As a result, we derive a measure of non-Markovianity for an incoherent open dynamical system as follows,
\begin{equation}
	\mathcal{N}^{CKD}\left( \Phi_{t}\right) = \max_{\varrho(0) \in \mathcal{I_{C}}} \int_{\sigma_{\mathcal{C}}\left( t\right)>0} \sigma_{\mathcal{C}}\left( t\right) dt,  \label{NKD}
\end{equation}
where the set of all coherent states is $\mathcal{I_{C}}$. Only for incoherent dynamics may the measure $\mathcal{N}^{CKD}\left( \Phi_{t}\right)$ see the non-Markovian feature. However, for certain particular coherence measures, it does not require auxiliary systems to make computations easier. Therefore, $\mathcal{N}^{CKD}\left( \Phi_{t}\right)$ is a positive function associated with the $\Phi$ family of dynamical maps, which is employed to quantify the maximum flow of information returning from the environment to the open system. By definition, this function cancels out, i.e. $\mathcal{N}^{CKD}\left(\Phi_{t}\right)=0$, only if the process is Markovian. Please note that growth in coherence via KD quasiprobability is a necessary but not sufficient condition for non-Markovianity.
\section{Applications} \label{sec4}
This section presents applications of our measure for evaluating non-Markovian processes in single- and two-qubit model systems. By analyzing these cases, we demonstrate how our measure can be used to effectively characterize quantum dynamical behavior.
\subsection{Non-Markovianity based on KD quasiprobability coherence in Single-qubit}
\subsubsection{Dephasing channel for a single qubit}

Here, we focus on a single-qubit system subjected to a dephasing channel. The Hamiltonian that describes the interaction with a thermal reservoir \cite{Breuer2002}, is written as follows
\begin{equation}
H=\omega_{0}\sigma_{z}+\sum\limits_{k}\omega_{k}a_{k}^{\dagger}a_{k}+\sum\limits_{k}\left( g_{k}\sigma_{z}a_{k}+g_{k}^{*}\sigma_{z}a_{k}^{\dagger}\right),
\end{equation}
For this model, the qubit's transition frequency is denoted by $\omega_{0}$, the Pauli-$z$ operator by $\sigma_{z}$, and the annihilation (creation) operator for the $k$-th reservoir mode with frequency $\omega_{k}$ by $a_{k}(a_{k}^{\dagger})$. The coupling strength between the qubit and the $k$-th reservoir mode is given by $g_{k}$. The master equation describing the temporal evolution of the system is written by
\begin{equation}
	\dot{\varrho}\left( t\right) = \gamma\left( t\right) \left[ \sigma_{z} \varrho\left( t\right) \sigma_{z}-\varrho\left( t\right) \right]. \label{master1}
\end{equation}
For initial state $\varrho(0)=\begin{pmatrix}
\varrho_{11}(0)   &&\varrho_{12}(0) 
\\\varrho_{21}(0) && \varrho_{22}(0)
\end{pmatrix}$, we have the dynamical map $(\varrho(t)=\Phi_{t}\varrho(0))$ for the dephasing channel as
\begin{equation}
	\varrho\left(t\right)=\begin{pmatrix}
	\varrho_{11}(0)  &&  \mathcal{R}\left(t\right) \varrho_{12}(0)
	\\ \mathcal{R}\left(t\right) \varrho_{21}(0) && \varrho_{22}(0)
	\end{pmatrix},
\end{equation}
with $\mathcal{R}\left(t\right) =e^{ -2\zeta\left(t\right)}$ with $\zeta\left(t\right) =\int_{0}^{t} \gamma\left( v\right) dv$.\par 
Let consider two eigenstates for observables $O_{\mu}$ and $O_{\nu}$ are $\left\lbrace \ket{\mu}\right\rbrace = \left\lbrace \ket{\mu_{0}}, \ket{\mu_{1}}\right\rbrace$ and $\left\lbrace \ket{\nu}\right\rbrace = \left\lbrace \ket{\nu_{+}}, \ket{\nu_{-}}\right\rbrace$ with

\begin{align}
    &\ket{\mu_{0}} = \ket{0}; \quad \ket{\nu_{+}} = \frac{1}{\sqrt{2}}(\ket{0}+i\ket{1}),\notag\\
    &\ket{\mu_{1}} = \ket{1}; \quad \ket{\nu_{-}}= \frac{1}{\sqrt{2}}(\ket{0}-i\ket{1}), \label{base1}
\end{align}
For the initial state 
\begin{equation}
    \varrho(0)=\ket{\phi}\bra{\phi} \label{rho01}
\end{equation}
with $\ket{\phi}=\frac{1}{\sqrt{2}}(\ket{0}+\ket{1})$, we have to calculate the KD quasiprobability $\mathcal{P}_{KD}$ by using eq.(\ref{P}) and the nonclassicality $N_{c}$ by using eq.(\ref{nonclass}) in the evolved state $\varrho(t)$,
\begin{equation}
    \mathcal{P}_{KD}(\mu,\nu|\varrho(t))= \frac{1}{2}(1+i\mathcal{R}(t)),
\end{equation}
\begin{equation}
    N_{c}(\mu,\nu|\varrho(t))= \frac{1}{2}(1+\mathcal{R}(t)).
\end{equation}
Then, can be compute CKD quasiprobability, $\mathcal{C}_{KD}[\varrho(t)]$ straightforwardly to give
\begin{equation}
    \mathcal{C}_{KD}[\varrho(t)]=\frac{1}{2}\mathcal{R}(t) 
\end{equation}
The temporal derivative of $\mathcal{C}_{KD}(t)$ takes the form
\begin{equation}
    \sigma_{c}(t) = -\gamma(t) \mathcal{R}(t),
\end{equation}
Consequently, the condition $\sigma_{c}(t) > 0$ is equivalent to $\gamma(t) < 0$, indicating non-Markovian behavior. The non-Markovianity measure $\mathcal{N}^{CKD}(\Phi_{t})$, associated with the $\Phi_{t}$ evolution, can then be expressed as follows
\begin{equation}
    \mathcal{N}^{CKD}(\Phi_{t}) = -\int_{\gamma(t)<0}\gamma(t)\mathcal{R}(t) dt
\end{equation}
Notably, although the measure based on the $\ell_{1}$-norm coherence detects non-Markovianity in a manner equivalent to the $\gamma(t)<0$ condition, this measure is not identical to the coherence measure of non-Markovianity. It has the following form \cite{Chanda2016},
\begin{equation}
    \mathcal{N}^{C_{l_{1}}}(\Phi_{t}) = -\int_{\gamma(t)<0} 2\gamma(t) \mathcal{R}(t) dt
\end{equation}
We examine a reservoir spectral density defined by the expression 
\begin{equation}
    \mathcal{S}_{d}(\omega) = (\frac{\omega}{\omega_{c}})^{s}\exp{(-\frac{\omega}{\omega_{c}})}, \label{J1}
\end{equation}
Here, $\omega_c$ denotes the reservoir cutoff frequency, and $s$ is the ohmicity parameter, which characterizes the type of reservoir: sub-ohmic for $s < 1$, ohmic for $s = 1$, and super-ohmic for $s > 1$. An analytical expression can be found for $\gamma\left( t\right)$ at zero temperature
\begin{equation}
\gamma\left( t\right) =\frac{\omega_{c}\Gamma\left[ s\right] \sin\left( \arctan\left( \omega_{c}t\right)\right)}{\left( 1+\left( \omega_{c}t\right)^{2}\right)^{s/2} },
\end{equation}
where $\Gamma\left[ z\right]$ the Euler gamma function.\\
\begin{widetext}

\begin{figure}[hbtp]
			{{\begin{minipage}[b]{.5\linewidth}
						\centering
						\includegraphics[scale=0.55]{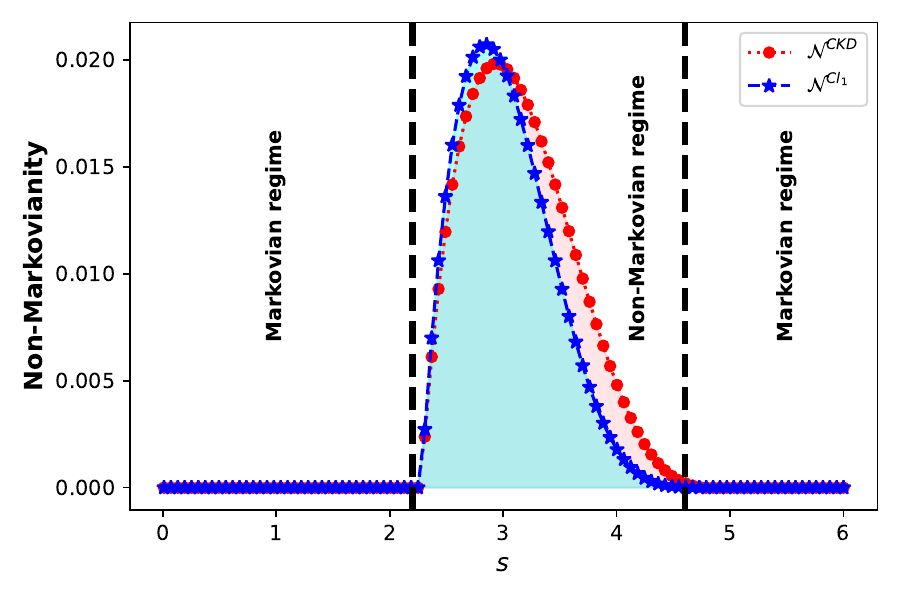} \vfill $\left(a\right)$
					\end{minipage}\hfill
					\begin{minipage}[b]{.5\linewidth}
						\centering
						\includegraphics[scale=0.6]{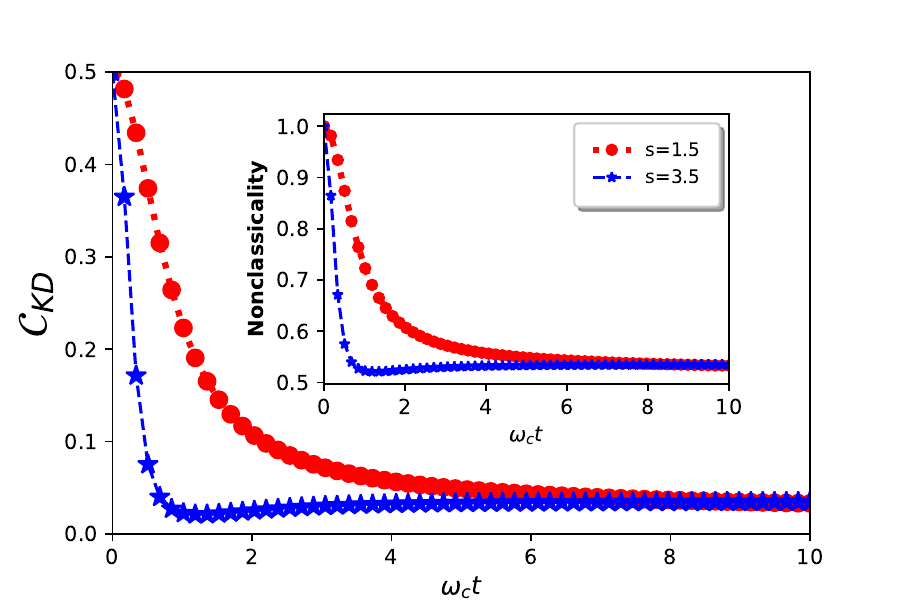} \vfill \vfill  $\left(b\right)$
			\end{minipage}}}
\caption{The variation of two non-Markovianity measures, $\mathcal{N}^{CKD}(\Phi_{t})$ based on the coherence via KD quasiprobability (Red-dashed), and $\mathcal{N}^{C_{l_{1}}}(\Phi_{t})$ based on the $\ell_{1}$-norm coherence (Blue-dashed) for a single-qubit dephasing channel with spectral density $\mathcal{S}_{d}(\omega)$ influenced by the ohmicity parameter $s$. For Fig. (b) we plot the behavior of the coherence via KD quasiprobability, $\mathcal{C}_{KD}$, and the nonclassicality $N_{c}$ measured by Eq. (\ref{nonclass}) (small plot) for the the dephasing channels, in the- scenarios, Markovian and non-Markovian regimes.}\label{Fig1}
\end{figure}
\end{widetext}

Fig.\ref{Fig1}(a) presents a comparative analysis of the evolution of the non-Markovian coherence measure $\mathcal{N}^{CKD}(\Phi_{t})$ and the measure $\mathcal{N}^{C_{l}}(\Phi_{t})$ as functions of the ohmicity parameter $s$. These two quantities become significant (non-zero) only for $s > 2.2$. We observe that $\mathcal{N}^{CKD}$ reaches a maximum around $s \simeq 3$, corresponding to the maximum information backflow in the system. Beyond this value, its amplitude gradually decreases due to the damping of oscillations in the dephasing rate $\gamma(t)$. Similarly, the $l_{1}$-norm coherence measure increases monotonically for $s > 2$, following the same behavior as our $\mathcal{N}^{CKD}(\Phi_{t})$ metric, indicating an increase in the frequency of regimes with a $\gamma(t)<0$. In Fig.\ref{Fig1}(b), we highlight the distinct behaviors of coherence via the KD quasiprobability $\mathcal{C}_{KD}$ and the nonclassicality dynamics $N_{c}(t)$ for a maximally coherent initial state, clearly distinguishing the Markovian and non-Markovian regimes.
\subsubsection{Single-qubit dissipative channel}
Now, we will examine the dissipative dynamics of a single qubit, using the following Hamiltonian, which is given by the formula
\begin{equation}
H=\omega_{0}\sigma^{+}\sigma^{-}+\sum_{k}\omega_{k}b_{k}^{\dagger}b_{k}+\sum\limits_{k}\left(g_{k}b_{k}\sigma^{+}+g_{i}^{*}b_{k}^{\dagger}\sigma^{-}\right) 
\end{equation}
where $\sigma^{+}\left( \sigma^{-}\right)$ represents the raising (lowering) Pauli operator, and $b_{k}$ and $b_{k}^{\dagger}$ is the creation and annihilation operators. It's associated dynamics is described using a master equation is
\begin{align}
   \dot{\varrho}\left( t\right)&=-\frac{i}{2}P\left( t\right)\left[ \sigma^{+}\sigma^{-},\varrho\left( t\right) \right] \notag\\& +\gamma\left( t\right) \Big( \sigma^{-}\varrho\left( t\right) \sigma^{+}-\frac{1}{2}[\sigma^{+}\sigma^{-},\varrho\left( t\right) ]_{+} \Big),
\end{align}
with the expression of $P\left( t\right)$ and $\gamma\left( t\right)$ is represented by
\begin{align}
	&P\left( t\right)=-2\Im\Big[\frac{\dot{\mathcal{B}}\left( t\right)}{\mathcal{B}\left( t\right)}\Big],\\
    &\gamma\left( t\right) =-2\Re\Big[\frac{\dot{\mathcal{B}}\left( t\right)}{\mathcal{B}\left( t\right)}\Big]=-\frac{2}{\arrowvert \mathcal{B}\left( t\right) \arrowvert}\frac{d}{dt}\arrowvert \mathcal{B}\left( t\right) \arrowvert,
\end{align}
where $\mathcal{B}\left( t\right)$ check the following integro-differential equation with the initial condition, 
\begin{align}
&\mathcal{B}\left( 0\right)=1 \notag\\
&\dot{\mathcal{B}}\left( t\right) = -\int_{0}^{t}G\left( t-t_{1}\right) \mathcal{B}\left( t_{1}\right) dt_{1} \label{R}
\end{align}
where the Fourier transform of the spectral density of the reservoir is the two-point correlation function of the reservoir $G\left( t-t_{1}\right)$,
\begin{equation}
	G\left( t-t_{1}\right)=\int  \mathcal{S}_{d}\left( \omega\right) \exp\left[ i\left( \omega_{0}-\omega\right) \left( t-t_{1}\right)\right]d\omega, \label{f}
\end{equation}
with $\mathcal{S}_{d}\left( \omega\right)$ is related to the spectral density. The Kraus representation $\varrho(t) = \sum_{n=1}^{2}\mathcal{K}_{n}(t)\varrho(0)\mathcal{K}_{n}^{\dagger}(t)$ the amplitude damping channel is given by
\begin{equation}
    \mathcal{K}_{1}(t)=\begin{pmatrix}
	1  &&  0
	\\ 0 && \mathcal{B}\left( t\right)
	\end{pmatrix}; \quad 
\mathcal{K}_{2}(t)=\begin{pmatrix}
	0  &&  \sqrt{1-|\mathcal{B}\left( t\right)|^{2}}
	\\ 0 && 0
	\end{pmatrix} 
\end{equation}
The evolution of the system qubit $\Phi_{t}\varrho\left( 0\right)$ for the initial state $\varrho(0)$ is written by
\begin{equation}
	\varrho\left( t\right) = \begin{pmatrix}
	1-\arrowvert \mathcal{B}\left( t\right) \arrowvert^{2}\varrho_{22}(0)   && \mathcal{B}^{*}\left( t\right)\varrho_{12}(0) 
	\\\mathcal{B}\left( t\right)\varrho_{21}(0) && \arrowvert \mathcal{B}\left( t\right) \arrowvert^{2}\varrho_{22}(0)
\end{pmatrix}.
\end{equation}
Considering that the structured reservoir spectral density exhibits a Lorentzian distribution \cite{Breuer2002}, i.e,
\begin{equation}
    \mathcal{S}_{d}\left( \omega\right) = \frac{\gamma_{0}\kappa^{2}}{2\pi\left( \left( \omega-\omega_{c}\right)^{2}+\kappa^{2}\right)}, \label{Sd2}
\end{equation}
where $\gamma_{0}$ is the coupling constant between the system and the reservoir, $\kappa$ is the width of the Lorentzian and $\omega_{c}$ is the central frequency of the distribution. This constant is the inverse of the relaxation time $\left(\tau_{s} = \frac{1}{\gamma_{0}}\right)$ and is linked to the Markovian decay of the system. The spectral width $\kappa$ is related to the reservoir and is the inverse of the correlation time $\left(\tau_{s} = \frac{1}{\kappa}\right)$. The dynamics of the system is considered to be Markovian in the weak coupling regime, characterized by $\gamma_{0} < \frac{\kappa}{2}$ \cite{Dhar2015}. In contrast, the development is non-Markovian if $\gamma_{0} > \frac{\kappa}{2}$. The following is obtained by using the Lorentzian spectral density and the function $\mathcal{B}\left( t\right)$,
\begin{equation}
	\mathcal{B}\left( t\right)=e^{\frac{-\left( \kappa-i\varpi\right)t}{2}}\left[  \cosh\left( \frac{\Delta t}{2}\right) +\frac{\kappa-i\varpi}{2}\sinh\left( \frac{\Delta t}{2}\right) \right],
\end{equation}
where $\Delta=\sqrt{\left( \kappa-i\varpi\right)^{2}-2\gamma_{0}\kappa}$ and $\left( \varpi=\omega_{0}-\omega_{c}\right)$  is the system-reservoir frequency detuning, with $\omega_{0}$ the qubit frequency.\par

For initial state (\ref{rho01}) $\varrho(0)$, the KD quasiprobability and nonclassicality expressions can be expressed simply as
\begin{align}
    &\mathcal{P}_{KD}(\mu,\nu|\varrho(t)) = \frac{1}{4}[2+i(\mathcal{B}(t)+\mathcal{B}^{*}(t))], \label{P3}\\
    &N_{c}(\mu,\nu|\varrho(t)) = \frac{1}{4}[1+|\mathcal{B}(t)|]. 
\end{align}
Therefore, by using Eq. (\ref{P3}) the expression of CKD quasiprobability is given by
\begin{equation}
    \mathcal{C}_{KD}[\varrho(t)] =  \frac{1}{2}|\mathcal{B}(t)|
\end{equation}
After some algebraic calculations, we then conclude that the degree of non-Markovianity $\mathcal{N}^{CKD}(\Phi_{t})$ for the channel, is expressed by
\begin{equation}
    \mathcal{N}^{CKD}(\Phi_{t}) = \frac{1}{2}\int_{\frac{d}{dt}|\mathcal{B}(t)|>0} \frac{d}{dt}|\mathcal{B}(t)|dt
\end{equation}
\begin{widetext}

\begin{figure}[hbtp]
			{{\begin{minipage}[b]{.5\linewidth}
						\centering
						\includegraphics[scale=0.56]{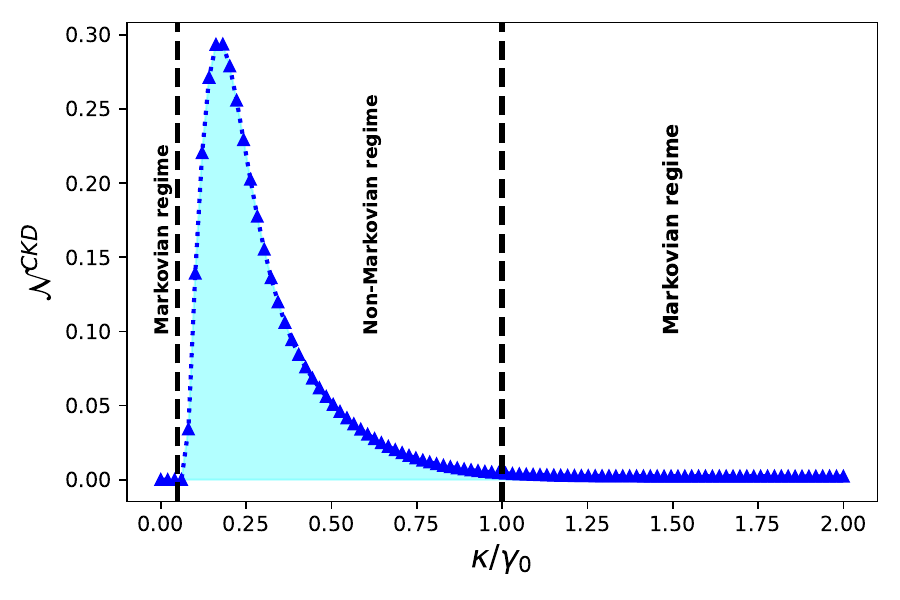} \vfill $\left(a\right)$
					\end{minipage}\hfill
					\begin{minipage}[b]{.5\linewidth}
						\centering
						\includegraphics[scale=0.6]{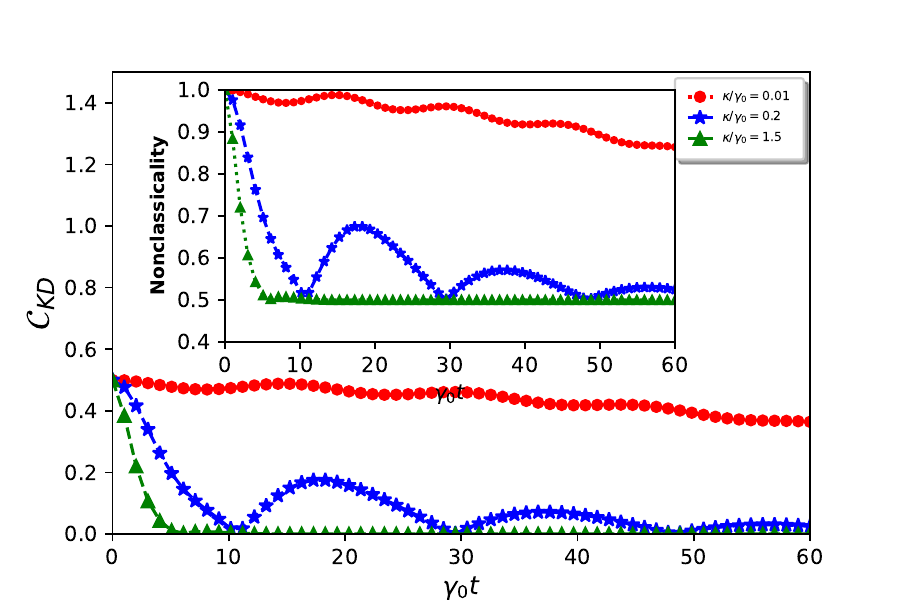} \vfill \vfill  $\left(b\right)$
			\end{minipage}}}
\caption{In Fig. (a) we plot coherence KD quasiprobability measurement of the non-Markovianity $\mathcal{N}^{CKD}(\Phi_{t})$ for a single-qubit dissipative channel, with a Lorentzian spectral density $\mathcal{S}_{d}(\omega)$ given in eq. (\ref{Sd2}). We show the evolution of $\mathcal{N}^{CKD}(\Phi_{t})$ as a function of the ratio of the spectral width of the distribution and the system-reservoir coupling $\kappa/\gamma_{0}$. In the Markovian region $\kappa/\gamma_{0} \in [0, 0.05]$ and for $\kappa/\gamma_{0} > 1$. $\mathcal{N}^{CKD}= 0$, while in the non-Markovian region $\kappa/\gamma_{0} \in [0.05, 1]$, $\mathcal{N}^{CKD}(\Phi_{t})$ takes non-zero values. Fig. (b): Time evolution of the coherence via KD quasiprobability $\mathcal{C}_{KD}$ and of nonclassicality $N_{c}$ (small plot) in the Markovian ($\kappa/\gamma_{0} = 0.01$ and $\kappa/\gamma_{0} > 1$) and non-Markovian ($\kappa/\gamma_{0} = 0.2$) regimes. In the non-Markovian regime, a nonmonotonic behavior of nonclassicality and $\mathcal{C}_{KD}$ is observed, while in the Markovian regime, $\mathcal{C}_{KD}$ remains a monotonic decreasing function of time. The observation is the same in terms of comparing nonclassicality and CKD quasiprobability in a quantitative context.}\label{Fig2}
\end{figure}

\end{widetext}
Fig.\ref{Fig2}(a) shows the transition between Markovian and non-Markovian dynamics as a function of the ratio $\gamma_{0}/\lambda$, where $\gamma_{0}$ represents the system-environment coupling and $\lambda$ the spectral width of the reservoir. We observe that the non-Markovianity measure $\mathcal{N}^{CKD}(\Phi_{t})$ becomes non-zero only when $\gamma_{0}/\lambda>0.5$, marking a clear boundary between the two regimes. This critical threshold suggests that non-Markovianity emerges when interaction with the environment dominates relaxation effects, allowing reversible information exchange. Furthermore, Fig. \ref{Fig2}(b) shows the effect of this transition on the temporal dynamics of coherence $\mathcal{C}_{KD}$ and nonclassicality $N_{c}$ for a maximally coherent initial state. In the Markovian case, coherence decreases irreversibly, in accordance with a decoherence process. In the non-Markovian regime, however, the $\mathcal{C}_{KD}$ and $N_{c}$ exhibits oscillations and partial recoveries, the signature of a bidirectional information flow between the system and its environment.
\subsection{Non-Markovianity based on KD quasiprobability coherence in two-qubit}
\subsubsection{Dephasing channel with global reservoir for two qubit}
The Hamiltonian that describes two qubits interacting with each other and connected to a shared thermal reservoir is expressed as follows
\begin{equation}
H=H_{S}+\sum_{j}\omega_{j}a_{j}^{\dagger}a_{j}+\frac{s_{z}}{2}\sum_{j}(g_{j}a_{j}+g_{j}^{*}a_{j}^{\dagger})
\end{equation}
The Hamiltonian of the system $H_{S}$, is expressed as follows 
\begin{equation}
    H_{S}=\sum_{i=1}^{2} \frac{h_{i}}{2}\sigma_{i}^{z}+\lambda\sigma_{1}^{z}\sigma_{2}^{z},
\end{equation}
with $s_{z} = (\sigma_{1}^{z} + \sigma_{2}^{z})$ representing the total spin operator along the $z$ axis, and $\lambda$ being the strength coupling parameter between two-qubit. The dynamics of the system is characterized by the master equation
\begin{equation}
    \dot{\varrho}\left( t\right)=-i[H_{S},\varrho(t)]+\gamma(t)
    (s_{z}\varrho(t)s_{z}-\frac{1}{2}[ s_{z}^{2},\varrho(t)]_{+},
\end{equation}
where $\gamma(t)$ denotes the time-dependent dephasing rate determined from the spectral density function $\mathcal{S}_{d}(\omega)$ given in eq. (\ref{J1}). For an arbitrary two-qubit state, $\varrho^{AB}(0)$, the element of the density matrix, $\varrho^{AB}(t)$, follows from
\begin{align}
    &\varrho_{11}(t)=\varrho_{11}(0), \quad \varrho_{22}(t)=\varrho_{22}(0), \notag\\
    &\varrho_{33}(t)=\varrho_{33}(0), \quad \varrho_{44}(t)=\varrho_{44}(0), \notag\\
    &\varrho_{14}(t)=\varrho_{14}(0)e^{-i(h_{1}+h_{2})t-8\zeta(t)} \notag\\
    &\varrho_{23}(t)=\varrho_{23}(0) e^{-i(h_{2}-h_{2})t-2\zeta(t)} \notag\\
    &\varrho_{12}(t)=\varrho_{12}(0) e^{-i(\lambda+h_{2})t-2\zeta(t)} \notag\\
    &\varrho_{13}(t)=\varrho_{13}(0) e^{-i(\lambda+h_{1})t-2\zeta(t)} \notag\\
    &\varrho_{24}(t)=\varrho_{24}(0) e^{i(\lambda-h_{1})t-2\zeta(t)} \notag\\
    &\varrho_{34}(t)=\varrho_{34}(0) e^{i(\lambda-h_{2})t-2\zeta(t)}, \label{matrix}
\end{align} 
Let us compute the CKD quasiprobability of the state relative to the incoherent orthonormal product basis $\left\lbrace \ket{\mu}\right\rbrace$ = $\left\lbrace \ket{00},\ket{01},\ket{10},\ket{11}\right\rbrace$, the second basis, $\left\lbrace \ket{\nu} \right\rbrace$, is expressed in terms of the parametrization of the Bloch sphere as follows:
\begin{align}
&\ket{\nu_{1\pm}} = \cos\left(\frac{\alpha_{1}}{2}\right)\ket{0} \pm e^{i\beta_{1}}\sin\left(\frac{\alpha_{1}}{2}\right)\ket{1}, \notag\\
&\ket{\nu_{2\pm}} = \sin\left(\frac{\alpha_{2}}{2}\right)\ket{0} \pm e^{i\beta_{2}}\cos\left(\frac{\alpha_{2}}{2}\right)\ket{1}.
\label{base2}
\end{align} 
Here, the parameters $(\alpha_1, \beta_1)$ and $(\alpha_2, \beta_2)$ are the spherical coordinates that define the orientations of the qubits on the Bloch sphere. The complete basis is then obtained as the tensor product of these one-qubit states :
\begin{align}
&\ket{\nu_{1+}, \nu_{2+}} = \ket{\nu_{1+}} \otimes \ket{\nu_{2+}}, \quad \ket{\nu_{1+}, \nu_{2-}} = \ket{\nu_{1+}} \otimes \ket{\nu_{2-}}, \notag\\
&\ket{\nu_{1-}, \nu_{2+}} = \ket{\nu_{1-}} \otimes \ket{\nu_{2+}}, \quad \ket{\nu_{1-}, \nu_{2-}} = \ket{\nu_{1-}} \otimes \ket{\nu_{2-}}.
\end{align}
The initial state $\varrho^{AB}(0)$ is given by
\begin{equation}
    \varrho^{AB}(0)=\ket{\phi}^{AB}\bra{\phi}, \quad \ket{\phi}=\frac{1}{\sqrt{2}}(\ket{00}+\ket{11}), \label{rho033}
\end{equation}
and if we take $\alpha_{1}=\alpha_{2}=\frac{\pi}{2}$ and $\beta_{1}=\beta_{2}=0$, the expression of the KD quasiprobability $\mathcal{P}_{KD}(\mu,\nu|\varrho^{AB}(t))$ and nonclassicality $N_{c}(\mu,\nu|\varrho^{AB}(t))$ for the density matrix $\varrho^{AB}(t)$ (\ref{matrix}) can be simplified as follows
\begin{align}
    &\mathcal{P}_{KD}(\mu,\nu|\varrho^{AB}(t))=  \frac{1}{4}[1-i\mathcal{R}(t)^{4}\sin((h_{1}+h_{2})t)],  \label{PKD}\\
    &N_{c}(\mu,\nu|\varrho^{AB}(t))=  \frac{1}{4}[1+|\mathcal{R}(t)^{4}\sin((h_{1}+h_{2})t)|].
\end{align} 
By using Eq. (\ref{PKD}) the expression of CKD quasiprobability is given by
\begin{align}
    \mathcal{C}_{KD}[\varrho^{AB}(t)]&=\frac{1}{4}\arrowvert  \mathcal{R}(t)^{4}\sin((h_{1}+h_{2})t) \arrowvert.
\end{align}
\begin{widetext}

\begin{figure}[hbtp]
			{{\begin{minipage}[b]{.5\linewidth}
						\centering
						\includegraphics[scale=0.6]{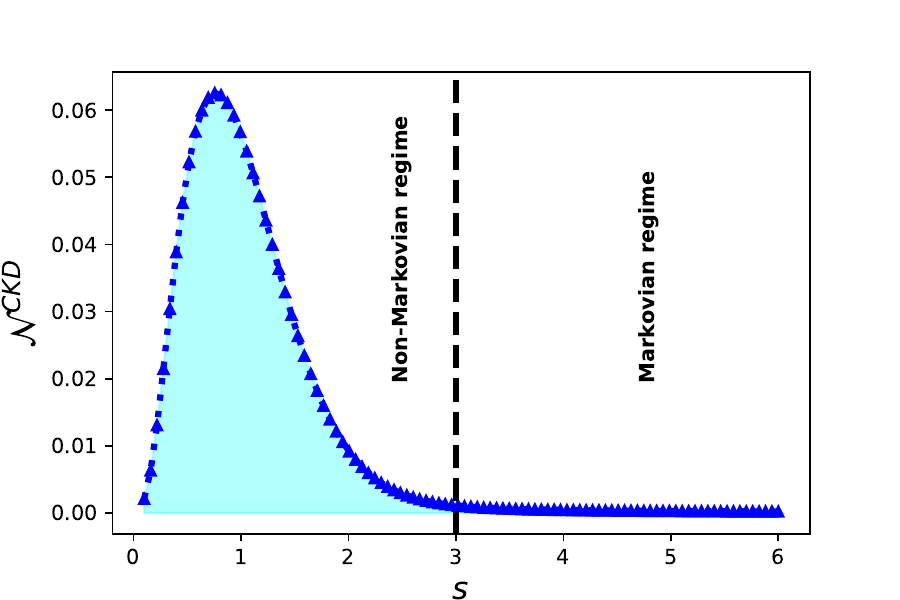} \vfill $\left(a\right)$
					\end{minipage}\hfill
					\begin{minipage}[b]{.5\linewidth}
						\centering
						\includegraphics[scale=0.6]{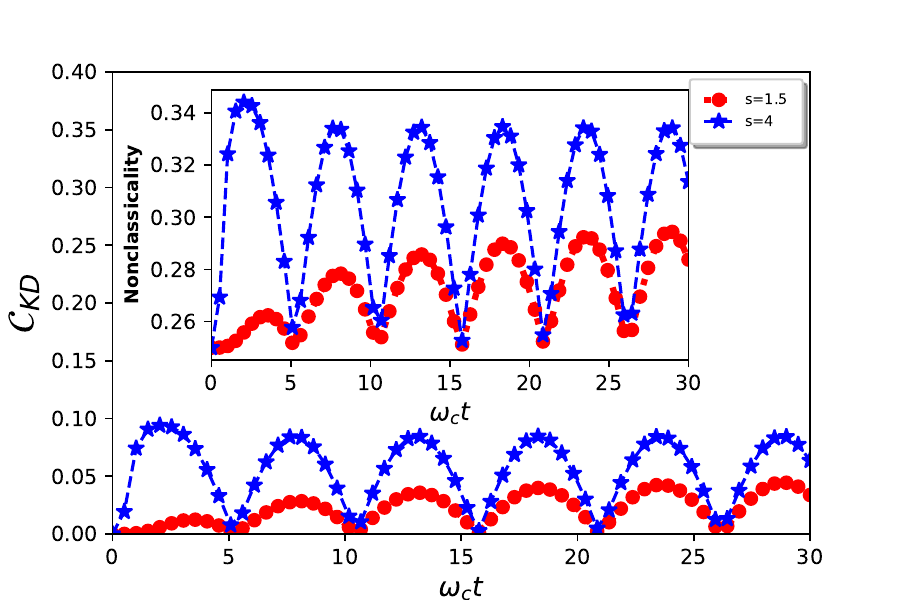} \vfill \vfill  $\left(b\right)$
			\end{minipage}}}
\caption{In Fig (a) we simplified quantum coherence based on KD quasiprobability of the non-Markovianity $\mathcal{N}^{CKD}(\Phi_{t})$ for dephasing channel a two-qubits, with an ohmic spectral density defined in eq. (\ref{J1}). We plot the variation of $\mathcal{N}^{CKD}(\Phi_{t})$ as a function of the Ohmicity parameter $s$. $\mathcal{N}^{CKD}(\Phi_{t})$ takes non-zero values in the interval $[0,3]$ in the non-Markovian regime and zero values for $s>3$ in the Markovian regime. For Fig.(b)  the comparison of the time evolution of quantum coherence based on KD quasiprobability and nonclassicality $N_{c}$ (small plot) for a two-qubit dephasing channel in the non-Markovian and Markovian regimes, with coupling constants $h_{1}=0.2$ and $h_{2}=0.4$.} \label{Fig3}
\end{figure}
\end{widetext}

Fig.\ref{Fig3}(a) illustrates the behavior of the simplified non-Markovianity measure $\mathcal{N}^{CKD}(\Phi_t)$ constructed from coherence using the KD quasiprobability as a function of the ohmicity parameter $s$. This parameter characterizes the spectral density structure of the environment and hence its memory properties. The figure shows that for $s < 3$ the measure $\mathcal{N}^{CKD}(\Phi_{t})$ is non-zero, signaling the presence of a backflow of information from the environment to the system, a clear signature of non-Markovian dynamics. As $s$ increases beyond this threshold, the measure cancels out, indicating that the environment is effectively memoryless and that the dynamics are shifting into the Markovian regime. Fig.\ref{Fig3}(b) completes this analysis by plotting the time evolution of coherence via KD quasiprobability, $\mathcal{C}_{KD}(t)$, as well as nonclassicality $N_{c}$ (see inset small plot) for two different values of coupling: $h_{1} = 0,2$ and $h_{2} = 0,4$. In the case of weak coupling (Markovian regime), $\mathcal{C}_{KD}(t)$ decreases smoothly and monotonically, indicating a continuous loss of quantum coherence without recovery. In contrast, for stronger coupling (non-Markovian regime), the function exhibits nonmonotonic behavior, with visible recoveries of coherence, physically correspond to a backflow of quantum information from the environment to the system. The small plot, representing the non-classicality $N_{c}$, shows similar dynamics, but with a generally higher amplitude.

\subsubsection{Amplitude damping channel for a two qubit}
For two-qubit amplitude damping channel, we consider the Hamiltonian

\begin{align}
H=&\sum_{i=A,B}\omega_{0}\sigma_{z}^{i}+\sum\limits_{k}\omega_{k}a_{k}^{\dagger}a_{k} \notag\\
    &+\sum_{k}(g_{k}a_{k}\sigma_{+}^{A}+g_{k}^{.}a_{k}^{\dagger}\sigma_{-}^{B})\sum_{k}(g_{k}a_{k}\sigma_{+}^{B}+g_{k}^{.}a_{k}^{\dagger}\sigma_{-}^{A})
\end{align}
After the system has undergone amplitude damping, the initial state $\varrho^{AB}(0)$ results in the diagonal and off-diagonal elements of the amplitude damping channel $\Phi_{t}^{AB}\varrho(0)$ being written as follows
\begin{align}
    &\varrho_{11}(t)= |\mathcal{B}(t)|^{4}\varrho_{11}(0), \notag\\
    &\varrho_{22}(t)= |\mathcal{B}(t)|^{2}\varrho_{11}(0)(1-|\mathcal{B}(t)|^{2})+\varrho_{22}(0)|\mathcal{B}(t)|^{2}, \notag\\
    &\varrho_{33}(t)= |\mathcal{B}(t)|^{2}\varrho_{11}(0)(1-|\mathcal{B}(t)|^{2})+\varrho_{33}(0)|\mathcal{B}(t)|^{2}, \notag\\
    &\varrho_{44}(t)=1-(\varrho_{11}(t)+\varrho_{22}(t)+\varrho_{33}(t)), \notag\\
    &\varrho_{12}(t)=|\mathcal{B}(t)|^{2} \mathcal{B}(t) \varrho_{12}(0); \quad \varrho_{13}(t)=|\mathcal{B}(t)|^{2} \mathcal{B}(t) \varrho_{13}(0), \notag\\
    &\varrho_{14}(t)=\mathcal{B}(t)^{2}\varrho_{14}(0), \quad \varrho_{23}(t)=|\mathcal{B}(t)|^{2}\varrho_{23}(0), \notag\\
    &\varrho_{24}(t)= \mathcal{B}(t)\varrho_{12}(0)(1-|\mathcal{B}(t)|^{2})+\varrho_{24}(0)\mathcal{B}(t), \notag\\
    &\varrho_{34}(t)= \mathcal{B}(t)\varrho_{12}(0)(1-|\mathcal{B}(t)|^{2})+\varrho_{34}(0)\mathcal{B}(t)
\end{align}
For initial state (\ref{rho033}), and by using eigenstates (\ref{base2}), the expression of KD quasiprobability is given by
\begin{align}
    \mathcal{P}_{KD}(\mu,\nu|\varrho^{AB}(t)) &= \cos^{2}(\frac{\alpha_{1}}{2})+\cos^{2}(\frac{\alpha_{2}}{2})+\frac{1}{8}\Big[\mathcal{B}^{2}(t)e^{i(\beta_{1}+\beta_{2})}\notag\\
    &-\mathcal{B}^{*}(t)^{2}e^{-i(\beta_{1}+\beta_{2})} \Big]\sin(\alpha_{1})\sin(\alpha_{2})
\end{align}
We take $\alpha_{1}=\alpha_{2}=\pi/2$ and $\beta_{1}=\beta_{2}=0$, the expression of KD quasiprobability $\mathcal{P}_{KD}$ and nonclassicality $N_{c}$ is
\begin{align}
    &\mathcal{P}_{KD}(\mu,\nu|\varrho^{AB}(t)) = \frac{1}{4}+\frac{1}{8}[\mathcal{B}(t)^{2}+\mathcal{B}^{*}(t)^{2}],\\
    &N_{c}(\mu,\nu|\varrho^{AB}(t)) = \frac{1}{4}(1+|\mathcal{B}(t)|^{2}).
\end{align}
So, the expression of CKD quasiprobability is given by
\begin{align}
    \mathcal{C}_{KD}[\varrho^{AB}(t)] =\frac{1}{4}|\mathcal{B}(t)|^{2}
\end{align}
\begin{widetext}

\begin{figure}[hbtp]
			{{\begin{minipage}[b]{.5\linewidth}
						\centering
						\includegraphics[scale=0.6]{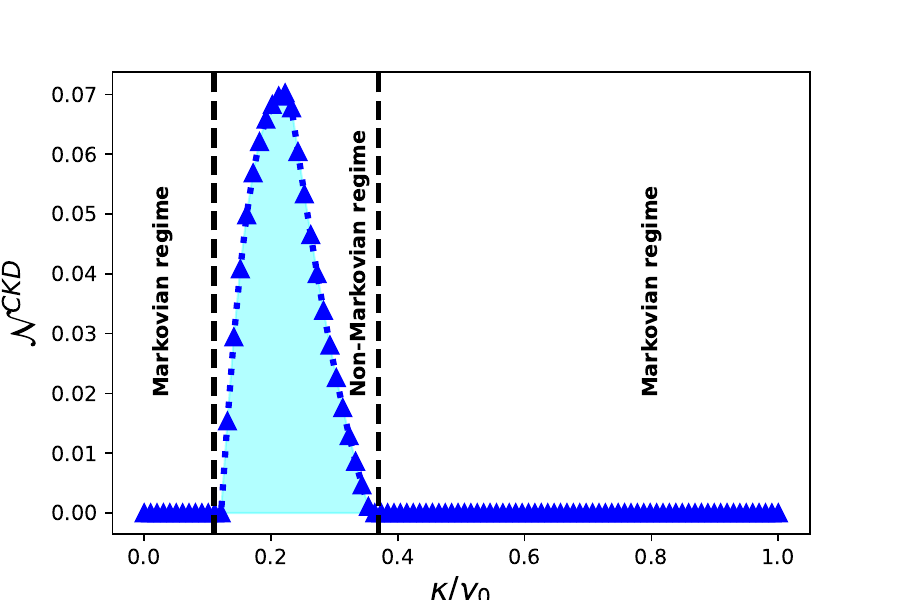} \vfill $\left(a\right)$
					\end{minipage}\hfill
					\begin{minipage}[b]{.5\linewidth}
						\centering
						\includegraphics[scale=0.6]{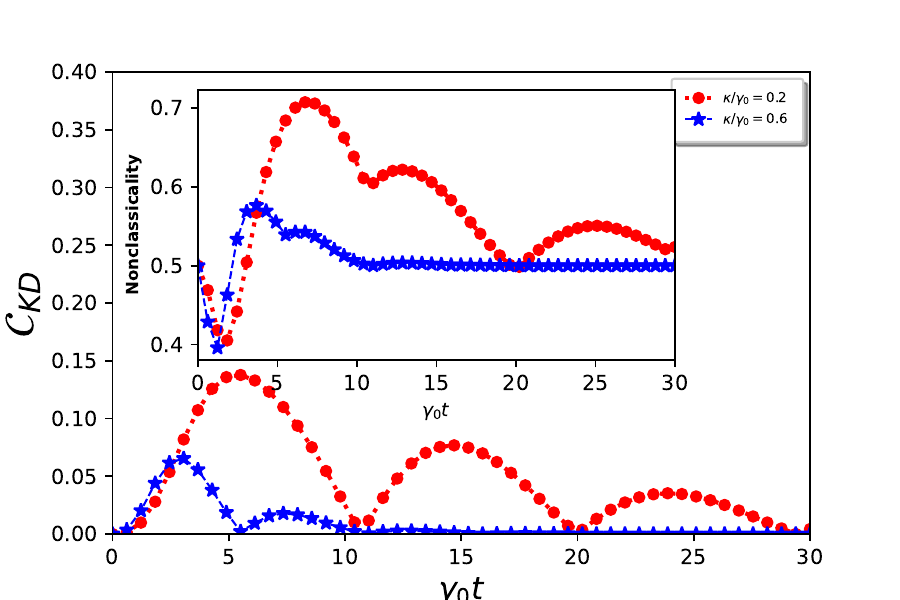} \vfill \vfill  $\left(b\right)$
			\end{minipage}}}
\caption{Fig. (a): Measurement of non-Markovianity for two-qubit amplitude damping channel, with the Lorentzian spectral density given by eq. (\ref{Sd2}). We plot $\mathcal{N}^{CKD}(\Phi_{t})$ as a function of the ratio of the spectral width of the distribution and the system-reservoir coupling $\kappa/\gamma_{0}$. The region $\kappa/\gamma_{0} \in [0, 0.1]$ and $\kappa/\gamma_{0} > 0.4$ indicates a Markovian regime, while the region $\kappa/\gamma_{0} \in [0.1, 0.35]$ corresponds to a non-Markovian regime. Fig. (b): The Behavior of the CKD quasiprobability and nonclassicality $N_{c}$ (small plot) in a two-qubit system evolving in an amplitude damping channel. We observe the evolution of the CKD quasiprobability and the non-classicality in the Markovian and non-Markovian regimes. In the Markovian regime $(\kappa/\gamma_{0} = 0.2)$, the CKD quasiprobability and nonclassicality decrease monotonically, while in the non-Markovian regime $(\kappa/\gamma_{0} = 0.6)$, a nonmonotonic behavior is observed.}\label{Fig5}
\end{figure}
\end{widetext}
Fig. \ref{Fig5} shows the analysis of non-Markovianity for two-qubit amplitude damping channel. In Fig. \ref{Fig5}(a), we study the evolution of the non-Markovianity measure $\mathcal{N}^{CKD}(\Phi_{t})$, based on quantum coherence via the imaginary part of KD quasiprobability, as a function of the ratio $\kappa/\gamma_{0}$, where $\kappa$ represents the spectral width of the environment (of Lorentzian form) and $\gamma_{0}$ the coupling between the system and the reservoir. This ratio controls the memory of the environment, a low $\kappa/\gamma_0$ corresponds to a long-memory reservoir, favoring non-Markovian behavior, while a high $\kappa/\gamma_0$ corresponds to a fast-response environment, typical of a Markovian regime. The results show that non-Markovianity is significant only in the intermediate range $\kappa/\gamma_{0} \in [0.1,0.35]$, characterized by information backflow effects. Outside this range, the system evolves in a Markovian manner, with an irreversible loss of information to the environment. For Fig. \ref{Fig5}(b) examines the dynamic evolution of $\mathcal{C}_{KD}$, as well as of nonclassicality $N_{c}$, for a two-qubit system subjected to a amplitude damping channel, which models energy losses such as spontaneous emission. In the Markovian regime $(\kappa/\gamma_{0} = 0.2)$, $\mathcal{C}_{KD}$ and $N_{c}$ decrease monotonically, illustrating a continuous and irreversible loss of quantum properties. In contrast, in the non-Markovian regime $(\kappa/\gamma_{0} = 0.6)$, we observe nonmonotonic behavior, with temporary revivals in $\mathcal{C}_{KD}$ and $N_{c}$, a clear sign of backflow from the environment to the system. This behavior confirms the ability of the coherence measure via KD quasiprobability, as well as the nonclassicality indicator, to finely detect memory effects in quantum dynamics.

\section{Conclusion} \label{clc}
This study introduces a novel measure of non-Markovianity for incoherent dynamical maps, based on quantum coherence as formulated via the Kirkwood–Dirac quasiprobability. The proposed approach is illustrated with prototypical examples that highlight its advantages and effectiveness compared to existing measures, such as those based on the $\ell_{1}$-norm coherence. We conduct a detailed study of dephasing and dissipative dynamics in single-qubit systems and provide a preliminary discussion of the two-qubit case, indicating that while our measure can coincide with $\ell_1$-norm coherence in certain scenarios, this equivalence does not generally hold, especially in systems with more complex correlations. Specifically, we track the transition between Markovian and non-Markovian regimes by analyzing the nonmonotonic behavior of coherence via KD quasiprobability. Our results reveal how signatures of non-Markovianity, such as the transient growth of quantum coherence, are captured by variations in coherence via KD quasiprobability. These findings underscore the central role of coherence via KD quasiprobability in characterizing complex quantum dynamics and open new avenues for robust experimental detection of non-Markovianity. Furthermore, our study provides clear evidence that this measure effectively captures key non-Markovian properties, such as information feedback from the environment to the system, while offering a potentially less demanding alternative in terms of computation, due to its reliance on a fixed coherence-based criterion derived from the KD quasiprobability. The ability of non-Markovianity to enhance nonclassicality in the presence of decoherence and decay highlights its potential for mitigating information loss, a fundamental challenge in modern quantum technologies. By offering a theoretical framework for harnessing this property, our work strengthens the tools available for practical quantum applications.\par

Finally, we hope this study will inspire further research to deepen our understanding of quantum dynamics in complex systems, whether biological, physical, or social. Such advances could pave the way for new approaches to simulating and controlling quantum systems in non-ideal environments, fostering the development of next-generation quantum technologies.\par

\end{document}